\documentclass[prX,twocolumn,nofootinbib]{revtex4}

\usepackage{graphicx}
\usepackage{amsmath,srcltx}
\usepackage{latexsym}
\usepackage{amssymb}    
\usepackage{amsfonts}

\usepackage{xcolor} 

\newcommand{\be}{\begin{equation}}
\newcommand{\ee}{\end{equation}}
\newcommand{\ba}{\begin{eqnarray}}
\newcommand{\ea}{\end{eqnarray}}

\def\a{\alpha}
\def\b{\beta}
\def\d{\delta}
\def\e{\epsilon}

\def\vf{\varphi}
\def\g{\gamma}
\def\h{\eta}
\def\j{\psi}

\def\m{\mu}
\def\n{\nu}
\def\w{\omega}
\def\p{\pi}

\def\r{\rho}
\def\s{\sigma}

\def\D{\Delta}
\def\F{\Phi}

\def\ca{{\cal A}}

\def\cj{{\cal J}}

\newcommand{\ov}{\overline}

\newcommand{\dg}{\dagger}
\newcommand{\aand}{\;\;\;\mbox{and}\;\;\;}
\newcommand{\pa}{\partial}

\newcommand{\sz}{\sigma_z}
\newcommand{\charg}{\stackrel{{C}}\longrightarrow}
\newcommand{\parit}{\stackrel{{P}}\longrightarrow}
\newcommand{\timer}{\stackrel{{T}}\longrightarrow}

\newcommand{\rae}{\rangle}
\newcommand{\lae}{\langle}

\newcommand{\oN}{{\mathbb N}}

\def\sl#1{\rlap{\hbox{$\mskip 1 mu /$}}#1}
\def\Sl#1{\rlap{\hbox{$\mskip 3 mu /$}}#1}
\def\SL#1{\rlap{\hbox{$\mskip 4 mu /$}}#1}
\def\I{\leavevmode\hbox{\small1\kern-3.8pt\norfeynmpmalsize1}}

\begin{document}
\title{On the electron-polaron--electron-polaron scattering and Landau levels in pristine graphene-like quantum electrodynamics}

\author{W.B. De Lima}
\email{wellissonblima@cbpf.br}
\affiliation{Centro Brasileiro de Pesquisas F\'isicas (CBPF),\\
Rua Dr. Xavier Sigaud 150 - 22290-180 - Urca - RJ - Brazil.}

\author{O.M. Del Cima}
\email{oswaldo.delcima@ufv.br}
\affiliation{Universidade Federal de Vi\c cosa (UFV),\\
Departamento de F\'\i sica - Campus Universit\'ario,\\
Avenida Peter Henry Rolfs s/n - 36570-900 - Vi\c cosa - MG - Brazil.}
\affiliation{Ibitipoca Institute of Physics (IbitiPhys),\\
36140-000 - Concei\c c\~ao do Ibitipoca - MG - Brazil.}

\author{E.S. Miranda}
\email{emerson.s.miranda@ufv.br}
\affiliation{Universidade Federal de Vi\c cosa (UFV),\\
Departamento de F\'\i sica - Campus Universit\'ario,\\
Avenida Peter Henry Rolfs s/n - 36570-900 - Vi\c cosa - MG - Brazil.}

\begin{abstract}
The parity-preserving $U(1)\times U(1)$ massless QED$_3$ is proposed as a pristine graphene-like planar quantum electrodynamics model. The spectrum content, the degrees of freedom, spin, masses and charges of the quasiparticles (electron-polaron, hole-polaron, photon and N\'eel quasiparticles) which emerge from the model are discussed. The four-fold broken degeneracy of the Landau levels, similar as the one experimentally observed in pristine graphene submitted to high applied external magnetic fields, is obtained. Furthermore, the model exhibits zero-energy Landau level indicating a kind of anomalous quantum Hall effect. The electron-polaron--electron-polaron scattering potentials in $s$- and $p$-wave states mediated by photon and N\'eel quasiparticles are computed and analyzed. Finally, the model foresees that two electron-polarons ($s$-wave state) belonging to inequivalent $\mathbf{K}$ and $\mathbf{K^\prime}$ points in the Brillouin zone might exhibit attractive interaction, while two electron-polarons ($p$-wave state) lying both either in $\mathbf{K}$ or in $\mathbf{K^\prime}$ points experience repulsive interaction. 
\end{abstract}
\maketitle

\section{Introduction}
The quantum electrodynamics in three dimensional space-time (QED$_3$) has drawn attention since the groundbreaking works by Schonfeld, Jackiw, Templeton and Deser \cite{deser-jackiw-templeton-schonfeld} owing to the viability of taking planar quantum electrodynamics models as theoretical foundation for quasiplanar condensed matter phenomena, such as high-$T_{\rm c}$ superconductors \cite{high-Tc}, quantum Hall effect \cite{quantum-hall-effect}, topological insulators \cite{topological-insulators}, topological superconductors \cite{topological-superconductors} and graphene \cite{graphene}. Since then, planar quantum electrodynamics models have been investigated in many physical arrangements, namely, small (perturbative) and large (non perturbative) gauge transformations, Abelian and non-Abelian gauge groups, fermions families, odd and even under parity, compact space-times, space-times with boundaries, curved space-times, discrete (lattice) space-times, external fields and finite temperatures. 

The pristine graphene, a monolayer of pure graphene \cite{graphene}, is a gapless quasibidimensional system behaving like a half-filling semimetal where the quasiparticles charge carriers are described by massless charged Dirac fermions. The electron-electron interactions in graphene \cite{electron-pairing} include electron-polarons \cite{electron-phonon} scattering processes \cite{polarons}, where the quasiparticle electron-polaron (or hole-polaron) is formed by a bound state of electron (or hole) and phonon \cite{landau}.

In this work a pristine graphene-like planar quantum electrodynamics model, the parity-preserving $U(1)\times U(1)$ massless QED$_3$, is proposed and introduced as follows. In Section II, the model defined by its discrete and continuous symmetries is presented and its spectrum -- degrees of freedom, spin, masses and charges of all the quanta particles content of the model -- is discussed. In graphene the interactions among the massless fermion quasiparticles (electron-polaron and hole-polaron) are nonconfining, so the vector meson mediated quasiparticles contained in the model, namely the photon and the N\'eel quasiparticles, must be massive -- massless mediated quanta in three space-time dimensions yield logarithm-type (confining) interaction potentials \cite{maris} -- consequently the asymptotic states for the massless fermion quasiparticles might be determined. Also, similar effect as the four-fold broken degeneracy of the Landau levels experimentally observed {\cite{landau-levels-exp} in pristine graphene under high applied external magnetic fields is noticed. Next, in Section III, the $s$- and $p$-wave M{\o}ller (electron-polaron--electron-polaron) scattering amplitudes are computed and their respective interaction potentials obtained and analyzed. Conclusions and final comments are left to Section IV.

\section{The model}
The proposed Lorentz invariant model for a pristine graphene-like planar quantum electrodynamics, the parity-even $U(1)_A\times U(1)_a$ massless QED$_3$, is defined by the action:
\ba
S&\!\!\!=\!\!\!&\int{d^3 x} \biggl\{i {\ov\j_+} {\Sl D} \j_+ + i {\ov\j_-} {\Sl D} \j_- +  \nonumber\\
&&~-{1\over4}F^{\m\n}F_{\m\n} - {1\over4}f^{\m\n}f_{\m\n} + \m\e^{\m\r\n}A_\m\pa_\r a_\n + \nonumber\\
&&~-\frac{1}{2\a}(\pa^\m A_\m)^2 - \frac{1}{2\b}(\pa^\m a_\m)^2 \biggr\}~,\label{action}
\ea
where ${\SL D}\j_\pm\equiv (\sl\pa + ie\Sl{A} \pm ig\sl{a})\j_\pm$, and any object 
${\SL X}\equiv X^\m\g_\m$. The coupling constants $e$ (electric charge) and $g$ (chiral charge) carry mass dimension $\frac12$, and the mixed Chern-Simons (CS) mass parameter $\m$ has mass dimension 1. The field strengths, $F_{\m\n}=\pa_\mu A_\nu - \pa_\n A_\m$ and $f_{\m\n}=\pa_\mu a_\nu - \pa_\n a_\m$, are associated to the electromagnetic field ($A_\m$) and the N\'eel (pseudo)chiral field ($a_\m$), respectively. The spinors $\j_+$ and $\j_-$ are two kinds of massless fermions, each of them representing electron-polaron (electron-phonon) and hole-polaron (hole-phonon) quasiparticles, where the subscripts $+$ (sublattice $\mathbf{A}$) and $-$ (sublattice $\mathbf{B}$) are related to their coupling to the N\'eel gauge field, or alternatively, to the two inequivalent $\mathbf{K}$ and $\mathbf{K^\prime}$ points in the Brillouin zone of a monolayer graphene. Also, the gamma matrices are fixed by $\g^\m=(\s_z,-i\s_x,i\s_y)$. It should be pointed out that possible parity-odd local counterterms as radiative corrections to the classical action (\ref{action}) might appear for the vacuum polarization tensor associated to the both gauge fields, $A_\m$ and $a_\m$, at 1-loop with linear ultraviolet degree of divergence ($\d=1$), and at 2-loops with logarithm ultraviolet degree of divergence ($\d=0$). In addition to that, at 1-loop parity-odd local counterterms might also arise for the self-energy related to the fermion fields, $\j_+$ and $\j_-$, with logarithm ultraviolet degree of divergence ($\d=0$). Nevertheless, by adopting the BPHZL (Bogoliubov-Parasiuk-Hepp-Zimmermann-Lowenstein) subtraction scheme and computing the contribution of those possible parity-violating local counterterms -- at 1- and 2-loops to the vacuum polarization tensor for the electromagnetic field ($A_\m$) and the N\'eel field ($a_\m$), as well as, at 1-loop to the self-energy for the sublattice $\mathbf{A}$ fermion ($\j_+$) and the sublattice $\mathbf{B}$ fermion ($\j_-$) -- the final conclusion is that they vanish \cite{1-loop-bphzl}. Furthermore, if parity symmetry could be broken or not -- at the stage of infrared subtractions \cite{masslessU1QED3,massiveU1QED3} induced by subtracting ultraviolet divergences during the BPHZL renormalization procedure at 1- and 2-loops orders, where potential ultraviolet divergences in vacuum polarization tensor and fermion self-energy Feynman graphs shall show up -- has also been verified. Also, it was demonstrated in \cite{1-loop-bphzl} that, neither parity-odd CS pure-like terms, $\e^{\m\r\n}A_\m\pa_\r A_\n$ or $\e^{\m\r\n}a_\m\pa_\r a_\n$, nor parity-even CS mixed-like term, $\e^{\m\r\n}A_\m\pa_\r a_\n$, just as neither fermion parity-odd monomials, $\ov\j_+\j_+$ or $\ov\j_-\j_-$, nor fermion parity-even binomial, $\ov\j_+\j_+ - \ov\j_-\j_-$, are radiatively generated. Finally, the action (\ref{action}) shows to be stable under quantum perturbation, thus multiplicative renormalizable, however it still lacks to prove its full renormalizability, namely the absence of any kind of anomaly at all orders in perturbation theory, which is now under investigation. 

\subsection{The symmetries: charge conjugation, parity, time reversion and $U(1) \times U(1)$}
The CPT-even action (\ref{action}) is invariant under the following discrete and continuous symmetries: 
\begin{enumerate}
\item charge conjugation symmetry ($C$):        
\ba
\j_\pm \!\! & \charg &\!\! \j_\pm^C=-\g^2\ov\j^{\top}_\pm~,~\ov\j_\pm \charg \ov\j_\pm^C=-\j_\pm^{\top}\g^2~,\nonumber \\
A_\mu \!\! & \charg &\!\! A_\mu^C=(-A_0,-A_1,-A_2)~,\nonumber\\
a_\mu \!\! & \charg &\!\! a_\mu^C=(-a_0,-a_1,-a_2)~. 
\label{xc}
\ea 
\item parity symmetry ($P$):        
\ba
x_\m \!\! & \parit &\!\! x_\m^P=(x_0,-x_1,x_2)~,\nonumber\\
\j_\pm \!\! & \parit &\!\! \j_\pm^P=-i\g^1\j_\mp~,~\ov\j_\pm \parit \ov\j_\pm^P=i\ov\j_\mp\g^1~,\nonumber \\
A_\mu \!\! & \parit &\!\! A_\mu^P=(A_0,-A_1,A_2)~,\nonumber\\
a_\mu \!\! & \parit &\!\! a_\mu^P=(-a_0,a_1,-a_2)~.
\label{xp}
\ea
\item time reversion symmetry ($T$):        
\ba
x_\m \!\! & \timer &\!\! x_\m^T=(-x_0,x_1,x_2)~,\nonumber\\
\j_\pm \!\! & \timer &\!\! \j_\pm^T=-i\g^2\j^{\ast}_\mp~,~\ov\j_\pm \timer \ov\j_\pm^T=i\ov\j^{\ast}_\mp\g^2~,\nonumber \\
A_\mu \!\! & \timer &\!\! A_\mu^T=(A_0,-A_1,-A_2)~,\nonumber\\
a_\mu \!\! & \timer &\!\! a_\mu^T=(-a_0,a_1,a_2)~. 
\label{xt}
\ea 
\item gauge $U(1)_A \times U(1)_a$ symmetry ($\delta_{\rm g}$):
\ba 
&&\delta_{\rm g} \psi_\pm(x)=i[\theta(x)\pm\omega(x)]\psi_\pm(x)~,~~\nonumber \\
&&\delta_{\rm g} \ov{\psi}_\pm(x)=-i[\theta(x) \pm \omega(x)]\ov{\psi}_\pm(x)~,\nonumber \\
&&\delta_{\rm g} A_{\mu}(x)=- \frac{1}{e}\,\partial_{\mu}\theta(x)~,\nonumber \\ 
&&\delta_{\rm g} a_{\mu}(x)=- \frac{1}{g}\,\partial_{\mu}\omega(x)~.\label{gaugetransf}
\ea
\end{enumerate}

\subsection{The spectrum: charges, spin, Landau levels, masses and degrees of freedom}
The free Dirac equations associated to massless spinors, $\j_+$ and $\j_-$, derived from the action (\ref{action}), read: 
\be
i{\sl\pa}\j_+=0 \aand i{\sl\pa}\j_-=0~.\label{dirac}
\ee
Thus, expanding the spinor field operators $\j_+$ and $\j_-$ in terms of the $c$-number plane wave solutions of the Dirac equations, with operator-valued amplitudes, $a_+$, $b_+$, $a_-$ and $b_-$ (annihilation operators), and $a_+^\dag$, $b_+^\dag$, $a_-^\dag$ and $b_-^\dag$ (creation operators), it follows that:
\ba
 \j_+(x)&=&\int\frac{d^2\vec p}{(2\p)\sqrt{2E}}\{a_+(p) u_+(p)e^{-ipx} + \nonumber\\ 
 &&+~ b_+^\dag(p) v_+(p)e^{ipx}\}~; \label{psi+}\\
 \j_-(x)&=&\int\frac{d^2\vec p}{(2\p)\sqrt{2E}}\{a_-(p) u_-(p)e^{-ipx} + \nonumber\\
 &&+~ b_-^\dag(p) v_-(p)e^{ipx}\}~, \label{psi-}
\ea
where $\ov\j_\pm=\j_\pm^\dag \g^0$. Consequently, taking into account (\ref{dirac}) and 
(\ref{psi+})-(\ref{psi-}), and by adopting $p^\m=(E,p_x,p_y)$ where $E=\sqrt{p_x^2 + p_y^2}$, since $p^\m p_\m=0$, the wave functions, $u_+$, $v_+$, $u_-$ and $v_-$, are given by:
\ba
u_+(p)=\frac{{\sl p}}{\sqrt{E}}u^{\prime}_+ ~,~~ u_-(p)=\frac{{\sl p}}{\sqrt{E}}u^{\prime}_- ~; \label{u+-}\\ 
v_+(p)=\frac{-{\sl p}}{\sqrt{E}}v^{\prime}_+ ~,~~ v_-(p)=\frac{-{\sl p}}{\sqrt{E}}v^{\prime}_- ~, \label{v+-} 
\ea 
fulfilling the conditions below:
\ba
&&u_+^\dg(p)u_+(p)=v_+^\dg(p)v_+(p)=2E ~; \label{norma+} \\ 
&&u_-^\dg(p)u_-(p)=v_-^\dg(p)v_-(p)=2E ~; \label{norma-} \\
&&\ov{u}_\pm(p)u_\pm(p)=\ov{v}_\pm(p)v_\pm(p)=0 ~, \label{norma}
\ea
where 
\ba 
&&u^{\prime}_+=u^{\prime}_-=\left(\begin{array}{c}
1  \\
0
\end{array}\right) 
\aand 
v^{\prime}_+=v^{\prime}_-=\left(\begin{array}{c}
0  \\
1
\end{array}\right) ~. \label{uvprime}
\ea  
 
From the microcausality conditions for the massless fermions $\j_+$ and $\j_-$: 
\be
\big\{\j_\pm(x),\j_\pm^\dag(y)\big\}_{x^0=y^0}=\d^2(\vec{x}-\vec{y})~,
\ee  
together with the Dirac equations (\ref{dirac}) and the normalization conditions 
(\ref{norma+})-(\ref{norma-}), it stems that:
\ba
&&\big\{a_\pm(p),a_\pm^\dag(k)\big\}=\d^2(\vec{p}-\vec{k})~,\\
&&\big\{b_\pm(p),b_\pm^\dag(k)\big\}= \d^2(\vec{p}-\vec{k})~,
\ea
where all other anticommutators vanish and, for the vacuum state $|0\rangle$, 
$a_\pm(k)|0\rangle=b_\pm(k)|0\rangle=0$.

\subsubsection{Charges}
The quantum operators associated with the internal $U(1)_A \times U(1)_a$ symmetry, namely electric charge ($Q_\pm$) and N\'eel (chiral) charge ($q_\pm$) operators, are 
\ba
Q_\pm \!\!&=&\!\! - e\int d^2\vec{x} :\j_\pm^{\dag}(x)\j_\pm(x): \label{Q+-}\\
      \!\!&=&\!\! - e\int d^2\vec p \{a_\pm^\dag(p) a_\pm(p) - 
      b_\pm^\dag(p) b_\pm(p)\} ~, \nonumber\\
q_\pm \!\!&=&\!\! \mp g\int d^2\vec{x} :\j_\pm^{\dag}(x)\j_\pm(x): \label{q+-}\\
      \!\!&=&\!\! \mp g\int d^2\vec p \{a_\pm^\dag(p) a_\pm(p) - 
      b_\pm^\dag(p) b_\pm(p)\} ~, \nonumber
\ea
respectively. Therefore, the electric charges and chiral charges of the asymptotic massless fermion (antifermion) states, $|f_{\uparrow}^-\rae$ ($|f_{\uparrow}^+\rae$) and $|f_{\downarrow}^-\rae$ 
($|f_{\downarrow}^+\rae$) read:  
\ba
&Q_+|f_{\uparrow}^-\rae = - e|f_{\uparrow}^-\rae ~,~~ 
Q_+|f_{\downarrow}^+\rae = + e|f_{\downarrow}^+\rae ~; \nonumber \\
&Q_-|f_{\downarrow}^-\rae = - e|f_{\downarrow}^-\rae ~,~~ 
Q_-|f_{\uparrow}^+\rae = + e|f_{\uparrow}^+\rae ~; \\
&q_+|f_{\uparrow}^-\rae = - g|f_{\uparrow}^-\rae ~,~~ 
q_+|f_{\downarrow}^+\rae = + g|f_{\downarrow}^+\rae ~; \nonumber \\
&q_-|f_{\downarrow}^-\rae = + g|f_{\downarrow}^-\rae ~,~~ 
q_-|f_{\uparrow}^+\rae = - g|f_{\uparrow}^+\rae ~;
\ea
where
\ba
|f_{\uparrow}^-\rae = a_+^\dag(k)|0\rae ~,&~~  |f_{\downarrow}^+\rae = b_+^\dag(k)|0\rae ~, \\ 
|f_{\downarrow}^-\rae = a_-^\dag(k)|0\rae ~,&~~ |f_{\uparrow}^+\rae = b_-^\dag(k)|0\rae ~, 
\ea
meaning that the creation operators $a_+^\dag$ and $a_-^\dag$ create a fermion (electron-polaron) whereas the creation operators $b_+^\dag$ and $b_-^\dag$ creates an antifermion (hole-polaron). The electron-polaron and hole-polaron electric and chiral charges are displayed in TAB. \ref{table1}.  

\subsubsection{(Pseudo)spin}
Whenever dealing with massless particles in three space-time dimensions, since there is no rest frame, spin  cannot be straightforwardly defined, nonetheless, spin (in the generalized sense of a quantum number labelling the representation of the little group \cite{binegar}) is still a fundamental quantum number. In the present case of massless fermions, it is verified that 
\ba
\left[H^{(0)}_{\pm}, L + \frac{1}{2}\sz \right]=0 ~, \label{spin}
\ea
where $H^{(0)}_{\pm}=\vec{\a}\cdot\vec{p}$ (with $\vec{\a}=\g^0\vec{\g}$) is the free Hamiltonian operator  associated to the massless spinors, $\j_+$ and $\j_-$, and $L=xp_y - yp_x$ is the angular momentum operator.   Thus, accordingly to (\ref{spin}) it shall be concluded that $\frac{1}{2}\sz$ is the (pseudo)spin operator. 

\subsubsection{Landau levels}
The issue of the (pseudo)spin can also be investigated by computing the quantum Landau levels of the model. Therefore, subjecting this pristine graphene-like system to high external and static magnetic field, $B=\frac{\partial A_y}{\partial x}-\frac{\partial A_x}{\partial y}$ and $A^\m=(0,\vec{A})$, somehow inducing within the bulk of the system a static magnetic-chiral field, $b=\frac{\partial a_y}{\partial x}-\frac{\partial a_x}{\partial y}$ and $a^\m=(0,\vec{a})$, the hamiltonians for the both massless spinors, $\j_+$ and $\j_-$, are respectively given by:
\ba
&&H_{+} = \vec{\alpha}\cdot(\vec{p}-e\vec{A}-g\vec{a})~; \label{HDCME+} \\
&&H_{-} = \vec{\alpha}\cdot(\vec{p}-e\vec{A}+g\vec{a})~, \label{HDCME-}
\ea   
whose spectrum, the quantum Landau levels, reads as follows:
\ba
&&E_{n,+,s} = \pm\sqrt{2(eB + gb)}\sqrt{n + \frac{1}{2} - (\underbrace{\pm \frac{1}{2}}_{\displaystyle s})}~,  \label{ENL+} \\
&&E_{n,-,s} = \pm\sqrt{2(eB - gb)}\sqrt{n + \frac{1}{2} - (\underbrace{\pm \frac{1}{2}}_{\displaystyle s})}~,  \label{ENL-}
\ea 
with $n$ being a non-negative integer number, $n \in \oN$ ($n=0,1,2,...$), where $E_{n,+,s}$ and $E_{n,-,s}$ are the Landau levels associated to $\j_+$ (at sublattice $\mathbf{A}$) and $\j_-$ (at sublattice $\mathbf{B}$) for electron-polarons (if $+$ sign in $E_{n,+,s}$ and $E_{n,-,s}$) or hole-polarons (if $-$ sign in $E_{n,+,s}$ and $E_{n,-,s}$) and $s=\pm\frac{1}{2}$ are the (pseudo)spin eigenvalues. The obtention of the spectrum for $g=0$ is well known \cite{graphene} and the calculation for the present case \cite{landau-levels-massless}\footnote{In \cite{landau-levels-massless}, the pristine graphene quantum electrodynamics model is introduced adopting the International System of Units, the quantum Landau levels computations are discussed in details, and the results compared with the experimental data of \cite{landau-levels-exp}.} is mathematically the same. The conceptual novelty is that the presence of the two types of fermions leads to two different cyclotron frequencies. This can be traced back to the difference in sign of the couplings of the two fermion flavors with the (pseudo)chiral field and possibly sheds light in the current debate \cite{landau-levels-exp} concerning the role played by spin and valley symmetries when graphene is submitted to a magnetic field. For example, it shall be stressed that the result displayed in (\ref{ENL+})-(\ref{ENL-}) mimics the four-fold broken degeneracy effect of the Landau levels (see FIG. \ref{landau_four_fold}) experimentally observed in pristine graphene under high applied magnetic fields \cite{landau-levels-exp}. Also, it yields from the equations (\ref{ENL+})-(\ref{ENL-}) that the lowest Landau level ($n=0$) appears at $E_{0,+,s}=E_{0,-,s}=0$ and accommodates electron-polarons or hole-polarons with only one pseudospin eigenvalue, namely $s=+\frac{1}{2}$, signalizing a possible anomalous-type quantum Hall effect. All other levels $n\geq1$ are occupied by electron-polarons or hole-polarons with both ($s=\pm\frac{1}{2}$) pseudospin eigenvalues. Therefore, this implies that for the lowest Landau level $n=0$ the degeneracy is half of that for any other $n\geq1$, likewise, all Landau levels ($n\geq1$) have the same degeneracy (a number of electron-polaron or hole-polaron states with a given energy) but the zero-energy ($n=0$) Landau level is shared equally by electron-polarons and hole-polarons, {\it i.e.} depending on the sign of the applied magnetic field there is only sublattice $\mathbf{A}$ or sublattice $\mathbf{B}$ states which contribute to the zero-energy (lowest) Landau level.

\begin{figure}[!t]
\centering
\setlength{\unitlength}{1,0mm}
\includegraphics[height=3.5cm]{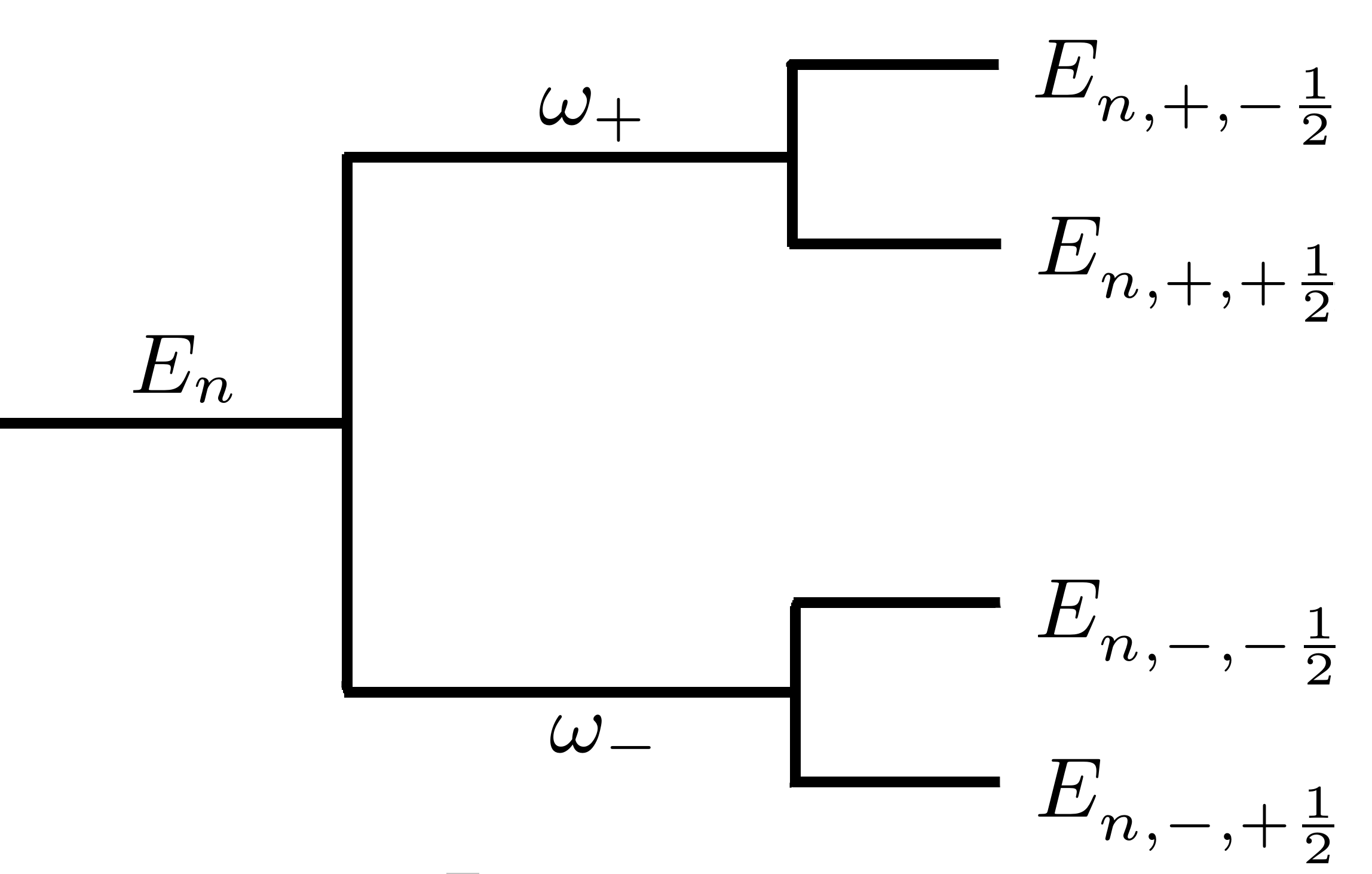}
\caption[]{Four-fold Landau levels of electron-polarons and hole-polarons at sublattices $\mathbf{A}$ and $\mathbf{B}$ with cyclotron frequencies $\w_+=\sqrt{2(eB + gb)}$ and $\w_-=\sqrt{2(eB - gb)}$, respectively, provided $n\geq1$.}
\label{landau_four_fold}
\end{figure}

An interesting issue comes to light, how the magnetic-chiral bulk-induced quantity ($gb$) could be measured in terms of physical quantities like external applied magnetic field ($B$), electron-polaron (hole-polaron) electric charge ($e$), pseudospin quantum number  ($s$) and Landau levels state energies ($E_{n,+,s}$ and $E_{n,-,s}$). One way would be by measuring the energy gaps of sublattices $\mathbf{A}$ ($\j_+$) and $\mathbf{B}$ ($\j_-$) electron-polarons (or hole-polarons) with pseudospin eigenvalue $s=+\frac{1}{2}$, from the first excited Landau state ($n=1$) to zero-energy state ($n=0$), 
$\D E_{10,+,+\frac{1}{2}} = E_{1,+,+\frac{1}{2}}-E_{0,+,+\frac{1}{2}}$ and $\D E_{10,-,+\frac{1}{2}} = E_{1,-,+\frac{1}{2}}-E_{0,-,+\frac{1}{2}}$, respectively, then the cyclotron frequencies $\w_+=\sqrt{2(eB + gb)}$ and $\w_-=\sqrt{2(eB - gb)}$ can be written as: 
\be
\w_+ = \mid\D E_{10,+,+\frac{1}{2}}\mid \aand \w_- = \mid\D E_{10,-,+\frac{1}{2}}\mid~,
\ee
and the bulk-induced quantity $gb$ reads:
\be
gb = \frac{(\D E_{10,+,+\frac{1}{2}})^2 - (\D E_{10,-,+\frac{1}{2}})^2}{4}~. \label{gb1}
\ee 
In the other way around, the bulk-induced quantity ($gb$) could be measured, for fixed Landau level quantum number ($n\geq1$) and pseudospin eigenvalue ($s$), by means of the sublattice $\mathbf{A}$ energy ($E_{n,+,s}$) and the sublattice $\mathbf{B}$ energy ($E_{n,-,s}$), such that from (\ref{ENL+})-(\ref{ENL-}), it yields that:
\be
gb = \frac{(E_{n,+,s})^2 - (E_{n,-,s})^2}{4(n+\frac{1}{2}-s)}~. \label{gb2}
\ee

\subsubsection{Masses and degrees of freedom}
For further computation on the electron-polaron--electron-polaron scattering amplitudes, the tree-level propagators in momenta space for all the fields have to be obtained and this can be achieved by switching off the coupling constants $e$ and $g$ in action (\ref{action}). Thus, it can be verified that: 
\ba
&&\D_{++}(k)=\D_{--}(k)=i\frac{\sl{k}}{k^2}~;\label{propk++--}\\
&&\D^{\m\n}_{AA}(k)=
-i\biggl\{ \frac{1}{k^2-\m^2}\biggl(\h^{\m\n}-\frac{k^\m k^\n}{k^2}\biggr) + 
\frac{\a}{k^2}\frac{k^\m k^\n}{k^2} \biggr\}~, \label{propkAA}\nonumber\\
&&\D^{\m\n}_{aa}(k)=
-i\biggl\{ \frac{1}{k^2-\m^2}\biggl(\h^{\m\n}-\frac{k^\m k^\n}{k^2}\biggr) + 
\frac{\b}{k^2}\frac{k^\m k^\n}{k^2} \biggr\}~, \label{propkaa}\nonumber\\
&&\D^{\m\n}_{Aa}(k)=\D^{\m\n}_{aA}(k)=\frac{\m}{k^2(k^2-\m^2)}\e^{\m\r\n}k_\r~. \label{propkAa}
\ea
From the propagators, $\D_{++}$ and $\D_{--}$ (\ref{propk++--}), $\D^{\m\n}_{AA}$, $\D^{\m\n}_{aa}$ and $\D^{\m\n}_{Aa}$ (\ref{propkAa}), the tree-level unitarity of the model, so as its spectrum, shall be verified by coupling 
them to conserved external currents  
$\cj_{\F_i}=(\cj_+,\cj_-,\cj^{\m}_A,\cj^{\m}_a)$ compatible with the symmetries of the model. Thereby, the current-current transition amplitudes in momentum space are written as $\ca_{\F_i\F_j} = \cj_{\F_i}^*(k) \lae \F_i(k) \F_j(k) \rae \cj_{\F_j}(k)$. Furthermore, picking up the imaginary part of the residues of the current-current amplitudes ($\ca_{\F_i\F_j}$) at the poles, it can be verified the necessary conditions for unitarity of the $S$-matrix at the tree-level -- positive imaginary part of the residues of the transition amplitudes, $\Im{\rm Res}~\ca_{\F_i\F_j}>0$ at all the poles -- besides the degrees of freedom counting of all the quantum fields presented in the model, $\F_i=(\j_+,\j_-,A_\m,a_\m)$. Briefly, it has been concluded \cite{msc-wellisson} that the two spinors, $\j_+$ and 
$\j_-$, hold two degrees of freedom -- the electron-polaron $|f_{\uparrow}^-\rae$ ($u_+$) and the hole-polaron $|f_{\downarrow}^+\rae$ ($v_+$) associated to the spinor $\j_+$, and the electron-polaron $|f_{\downarrow}^-\rae$ ($u_-$) and the hole-polaron $|f_{\uparrow}^+\rae$ ($v_-$) associated to the spinor $\j_-$. Moreover, the gauge fields, the electromagnetic field ($A_\m$) and 
the N\'eel field ($a_\m$), carry each one two massive degrees of freedom with mass $\m$. Also, the single massless mode in $\D^{\m\n}_{Aa}$ (\ref{propkAa}) presented in the interaction sector does not propagate, it decouples. Summarizing all results presented above, it is concluded that the parity-even $U(1)_A\times U(1)_a$ massless QED$_3$, a pristine graphene-like planar quantum electrodynamics model, is free from tachyons and ghosts at the classical level. Notwithstanding, to guarantee the unitarity at tree-level, it is still necessary to investigate the behaviour of the scattering cross sections by testing the fulfillment of the Froissart-Martin bound \cite{froissart-martin-bound} in the ultraviolet regime, as also in the infrared limit due to the presence of massless quantum fermions.

\begin{table}[b]
\begin{tabular}{|c|p{1,2cm}|p{1cm}|p{1cm}|c|} 
\hline
{state} & {$\;\,$ wave \newline function} & {electric charge} & {$\,$chiral charge} & {quasiparticle}  \\
\hline\hline
{$|f_{\uparrow}^-\rae$} & ~~~~$u_+$  & ~~{$-e$}   & ~~{$-g$}  & {electron-polaron} \\
\hline 
{$|f_{\downarrow}^-\rae$} & ~~~~$u_-$  & ~~{$-e$}   & ~~{$+g$}  & {electron-polaron} \\
\hline
{$|f_{\downarrow}^+\rae$} & ~~~~$v_+$  & ~~{$+e$}   & ~~{$+g$}  & {hole-polaron} \\
\hline
{$|f_{\uparrow}^+\rae$} & ~~~~$v_-$  & ~~{$+e$}   & ~~{$-g$}  & {hole-polaron} \\
\hline
\end{tabular}
\caption[]{The electron-polaron and hole-polaron electric and chiral charges.}\label{table1}
\end{table}

\section{The M{\o}ller scattering: electron-polaron--electron-polaron}
In order to calculate the scattering amplitudes, so as to obtain the scattering potentials, use has been made of the vertex Feynman rules associated to the interaction vertices $-e{\ov\j_\pm} \Sl{A} \j_\pm$ and 
$\mp g{\ov\j_\pm} \sl{a} \j_\pm$: $\Upsilon_{\pm \pm}^\m\!\!=\!\!ie\g^\m$ and 
$\upsilon_{\pm \pm}^\m\!\!=\!\!\pm ig\g^\m$, respectively. 

\begin{figure}[h]
\centering
\setlength{\unitlength}{1,0mm}
\includegraphics[width=5.5cm,height=2cm]{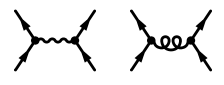}
\caption[]{$e^-$-polaron--$e^-$-polaron (M{\o}ller) $t$-channel scattering mediated  
by electromagnetic ($A_\mu$) and N\'eel ($a_\mu$) quantum fields.}
\label{moller}
\end{figure}

The $t$-channel in $s$- and $p$-wave states of electron-polaron--electron-polaron scattering amplitudes owing to electromagnetic and N\'eel quanta exchange (see FIG. \ref{moller}) are given by: 
\ba
&&\!\!\!\!\!\!\!\!-i\mathcal{M}_{\pm A \mp} = \nonumber\\
&&\!\!\!\!\!\!\!\!\ov{u}_\pm(p_1')[\Upsilon_{\pm\pm}^\m]u_\pm(p_1)\D_{\m\n}^{AA}(k)\ov{u}_\mp(p_2')[\Upsilon_{\mp\mp}^\n]u_\mp(p_2)~,\label{amplicovariante1}\\ 
&&\!\!\!\!\!\!\!\!-i\mathcal{M}_{\pm a \mp} = \nonumber\\
&&\!\!\!\!\!\!\!\!\ov{u}_\pm(p_1')[\upsilon_{\pm\pm}^\m]u_\pm(p_1)\D_{\m\n}^{aa}(k)\ov{u}_\mp(p_2')[\upsilon_{\mp\mp}^\n]u_\mp(p_2)~,\label{amplicovariante2}\\
&&\!\!\!\!\!\!\!\!-i\mathcal{M}_{\pm A \pm} = \nonumber\\
&&\!\!\!\!\!\!\!\!\ov{u}_\pm(p_1')[\Upsilon_{\pm\pm}^\m]u_\pm(p_1)\D_{\m\n}^{AA}(k)\ov{u}_\pm(p_2')[\Upsilon_{\pm\pm}^\n]u_\pm(p_2)~,\label{amplicovariante3}\\ 
&&\!\!\!\!\!\!\!\!-i\mathcal{M}_{\pm a \pm} = \nonumber\\
&&\!\!\!\!\!\!\!\!\ov{u}_\pm(p_1')[\upsilon_{\pm\pm}^\m]u_\pm(p_1)\D_{\m\n}^{aa}(k)\ov{u}_\pm(p_2')[\upsilon_{\pm\pm}^\n]u_\pm(p_2)~,\label{amplicovariante4}
\ea
with $\mathcal{M}_{\pm A \mp}$ and $\mathcal{M}_{\pm a \mp}$ being the scattering amplitudes in $s$-wave state, whereas $\mathcal{M}_{\pm A \pm}$ and $\mathcal{M}_{\pm a \pm}$ being those in $p$-wave state. 

The three-momenta configuration of the two scattered massless electron-polarons in the center of momenta (CM) reference frame; the incoming momenta, $p_1$ and $p_2$; the outgoing momenta, $p_1'$ and $p_2'$, as well as the momentum transfer, $k$, are defined below
\ba
&&\!\!\!\!\!\!\!\!\!\!\!\!p_1 = (E,p,0) ~,~~ p_1' = (E,p\cos\vf,p\sin\vf)~; \label{momenta1}\\
&&\!\!\!\!\!\!\!\!\!\!\!\!p_2 = (E,-p,0) ~,~~ p_2' = (E,-p\cos\vf,-p\sin\vf)~; \label{momenta2}\\
&&\!\!\!\!\!\!\!\!\!\!\!\!k = p_1 - p_1' = (0,p(1 - \cos\vf),-p \sin\vf) = (0, {\vec{k}})~, \label{momentak}
\ea 
where $\vf$ is the CM scattering angle, defined as the angle among the directions in the CM frame of the two incoming (initial state) and outgoing (final state) massless electron-polarons.  

Assuming the momenta configuration above (\ref{momenta1})-(\ref{momentak}) and taking into consideration the conditions (\ref{u+-})-(\ref{uvprime}) on the wave functions $u_+$ and $u_-$, the total $s$- and $p$-wave M{\o}ller scattering amplitudes can be obtained from the partial ones (\ref{amplicovariante1})-(\ref{amplicovariante4}), $\mathcal{M}_{\pm\mp}$ 
($\mid\uparrow\rae +\! \mid\downarrow\rae \rightarrow\;  \mid\uparrow\rae +\! \mid\downarrow\rae$) and 
$\mathcal{M}_{\pm\pm}$ 
($\mid\uparrow\rae +\! \mid\uparrow\rae \rightarrow\; \mid\uparrow\rae +\! \mid\uparrow\rae$ or 
$\mid\downarrow\rae +\! \mid\downarrow\rae \rightarrow\; \mid\downarrow\rae +\! \mid\downarrow\rae$), then it follows that:
\ba
\mathcal{M}_{\pm\mp}\!\!&=&\!\! -(e^2 - g^2) \left[\left(\frac{s-u}{t-\m^2}\right) -
\left(u\leftrightarrow t\right)\right]~; \label{Amplitude1}\\
\mathcal{M}_{\pm\pm}\!\!&=&\!\! (e^2 + g^2)e^{\pm i\vf}\left[\left(\frac{s-u}{t-\m^2}\right) + 
\left(u\leftrightarrow t\right)\right]~, \label{Amplitude2}
\ea
where, $s$, $t$ and $u$ are the Lorentz invariant Mandelstam variables evaluated at the CM frame:
\ba
s \!\!&=&\!\! (p_1+p_2)^2= 4E^2~, \nonumber\\
t \!\!&=&\!\! (p_1-p_1')^2=-2p^2(1-\cos\vf ) = -4p^2 \sin ^2\left(\frac{\vf}{2}\right)~, \nonumber\\
u \!\!&=&\!\! (p_1-p_2')^2=-2p^2(1+\cos\vf)=-4p^2\cos ^2\left(\frac{\vf}{2}\right)~. \nonumber 
\ea 

\subsection{The scattering potentials}
The two-particle interaction potential for two distinguishable electron-polarons (fermions), 1 and 2, in the tree approximation \cite{sucher} at the CM frame, reads:
\be
V(\vec{r}) = \int \frac{d^2\vec{k}}{(2\pi)^2} ~e^{i\vec{k}\cdot\vec{r}}\b_1 \b_2 F(\vec{k})~, \label{V}
\ee
with the product $\b_1\b_2$ being a spinorial factor in the space of the electron-polarons 1 and 2, where $\b_1=\g_1^0$ and $\b_2=\g_2^0$. Also, in addition to that, taking into account only the $t$-channel part of the electron-polaron--electron-polaron total scattering amplitude ($\mathcal{M}$) evaluated at the CM frame, it follows that  
\be
\mathcal{M} = \ov{u}_1(p_1') \ov{u}_2(p_2') F({k}) {u}_1(p_1) {u}_2(p_2)~. \label{M}
\ee
Moreover, in accordance to (\ref{amplicovariante1})-(\ref{amplicovariante4}), (\ref{V}) and (\ref{M}), the scattering potentials among two electron-polarons, firstly, in $s$-wave state -- one situated at $\mathbf{K}$ point and the other at $\mathbf{K^\prime}$ point in the Brillouin zone -- and secondly, in $p$-wave state -- the both located at either $\mathbf{K}$ point or $\mathbf{K^\prime}$ point in the Brillouin zone -- mediated by photon and N\'eel quasiparticles, can be respectively written as: 
\ba
&&V_{s}(r)= (1-\vec{\a}_1\cdot\vec{\a}_2)\frac{(e^2 - g^2)}{2\pi}K_0(\m r)~, \label{Vs}\\
&&V_{p}(r)= (1-\vec{\a}_1\cdot\vec{\a}_2)\frac{(e^2 + g^2)}{2\pi}K_0(\m r)~, \label{Vp}
\ea
where $\vec{\a}=\g^0\vec{\g}$. Thereafter, from (\ref{Vp}) it is concluded that, regardless the values of the electromagnetic ($e$) and the chiral coupling ($g$) constants, the electron-polaron--electron-polaron interaction in $p$-wave state ($|\mathbf{K}\rae + |\mathbf{K}\rae$ or $|\mathbf{K^\prime}\rae + 
|\mathbf{K^\prime}\rae$) is invariably repulsive. Nevertheless, drawing attention to (\ref{Vs}) it takes in evidence about the possibility of attractive electron-polaron--electron-polaron interaction in $s$-wave state ($|\mathbf{K}\rae + |\mathbf{K^\prime}\rae$) provided $g^2>e^2$. Notwithstanding, in spite of attractive potential be a necessary condition, although not a sufficient one, for the existence of $s$-wave ($\mathbf{K}$-$\mathbf{K^\prime}$) massless bipolarons bound states, it still remains to verify relativistic conditions similar as the ones whose were already verified for the non relativistic massive case \cite{mass-gap-qed3} -- where the non relativistic $s$-wave attractive scattering potential \cite{de_lima-del_cima-franco-neves} was proved to satisfy the Kato condition \cite{kato}, the Newton-Set\^o and the Bargmann upper bounds \cite{newton-seto,bargmann} -- which, in turn, guarantees that $s$-wave ($\mathbf{K}$-$\mathbf{K^\prime}$) massive bipolarons bound states exist. 

It shall be also mentioned that the presence of Breit-Darwin-type term $\vec{\a}_1\cdot\vec{\a}_2$ in (\ref{Vs}) and (\ref{Vp}), where in the non-zero gap (mass-gap) graphene-type \cite{mass-gap-qed3} is of order $\frac{v_1v_2}{c^2}$ (in real graphene units $\frac{v_1v_2}{v_F^2}$) and can be therefore neglected at low energies (low speeds), here in the pristine (gapless) graphene-type case, wherein the electron-polarons and hole-polarons are massless, cannot. In summary, it has yet to be investigated in relativistic regime if the attractive ($g^2>e^2$) $s$-wave state potential (\ref{Vs}) favours massless bipolarons, namely, electron-polaron--electron-polaron or hole-polaron--hole-polaron $s$-wave bound states. Furthermore, in order to avoid the continuum dissolution problem \cite{sucher}, the Dirac equation for the wave function ($\Psi$) associated to a two-particle-interaction system, $H_{\rm D}\Psi=E\Psi$, has to be rewritten as: 
\be
H_{\rm D}\Psi = H_{{\rm D}1}\Psi + H_{{\rm D}2}\Psi + V_{++}\Psi = E\Psi~,
\ee
where $V_{++}=\Lambda_{++}V\Lambda_{++}$, with $\Lambda_{++}=\Lambda(1)\Lambda(2)$ being the product of Casimir-type positive energy projection operators $\Lambda (i)=\frac{1}{2}\left({\mathbb I}+\frac{H_{{\rm D}i}}{E}\right)$ with $H_{{\rm D}i}=\vec{\a}_i\cdot\vec{p}_i$ ($i=1,2$). Besides that, bearing in mind that at the CM frame $\vec{p}_1=-\vec{p}_2\equiv\vec{p}$, thus 
\be
\vec{\a}_1\cdot\vec{p}\,\Psi - \vec{\a}_2\cdot\vec{p}\,\Psi + \Lambda_{++}V\Lambda_{++}\Psi = E\Psi~.\label{dirac-bound-state}
\ee
Finally, the question if whether or not, whenever $g^2>e^2$, the attractive $s$-wave state potential (\ref{Vs}) will favour $s$-wave massless bipolarons (two-fermion bound states) shall be definitively answered by investigating into details the Dirac equation (\ref{dirac-bound-state}), together with necessary and sufficient conditions \cite{brown-ravenhall-operator} which guarantee relativistic two-particle massless\footnote{There in \cite{brown-ravenhall-operator}, relativistic two-fermions bound states interacting via an attractive potential of the Bessel-Macdonald type -- the zeroth-order modified Bessel function of the second kind -- is analized for massive fermions. Therefore, for the massless case, a detailed investigation on this issue shall be pursued.} bound states. 

\section{Conclusions}
The model proposed, a gapless pristine graphene-like planar quantum electrodynamics model, the parity-preserving $U(1)\times U(1)$ massless QED$_3$, exhibits two-fermion scattering short range non confining potentials originated by two massive vector-mediated quanta, the photon (electric charge) and the N\'eel (chiral charge) quasiparticle, both stemming from the gauging of the $U(1)\times U(1)$ global symmetry. At the tree-level, the absence in the spectrum of tachyons ($k^2<0$) and ghosts ($\lae\j|\j\rae<0$) assures respectively, causality and unitarity, at this level. Additionally, the charges of the quasiparticles (electron-polaron, hole-polaron, photon and N\'eel quasiparticles), their masses, degrees of freedom and (pseudo)spin are determined and discussed. As a by-product, it is obtained the four-fold broken degeneracy of the Landau levels, reminding those experimentally observed in pristine graphene subjected to high external magnetic fields \cite{landau-levels-exp}, moreover, the system presents zero-energy Landau level suggesting a kind of anomalous quantum Hall effect -- detailing results and discussions shall appear further \cite{landau-levels-massless}. 

The $p$-wave state fermion--fermion (or antifermion--antifermion) scattering potential shows to be repulsive (\ref{Vp}) whatever the values of the electric ($e$) and chiral ($g$) charges. Nevertheless, for $s$-wave scattering of fermion--fermion (or antifermion--antifermion), the interaction potential (\ref{Vs}) might be attractive provided $g^2>e^2$. In summary, if two electron-polarons (or hole-polarons) lie in the inequivalent $\mathbf{K}$ and $\mathbf{K^\prime}$ points in the Brillouin zone the interaction might be  attractive, otherwise the interaction is always repulsive if those two electron-polarons (or hole-polarons) rest both either in $\mathbf{K}$ or in $\mathbf{K^\prime}$ points.

In view of possible applications of this quantum electrodynamics three space-time dimensional model to pristine (gapless) graphene, or any other planar system, the orders of magnitude of some theoretical parameters have to be estimated. The typical energy-scale in graphene -- for instance $E=v_F|\vec{p}|$, $C_s=\frac{1}{2\pi}(e^2-g^2)$ or $C_p=\frac{1}{2\pi}(e^2 + g^2)$ -- is around meV \cite{graphene}, while the length-scale interaction $\sl\lambda=\frac{2\p\hbar}{\m c}$ -- the reduced Compton wavelength of the quantum-mediated photon and N\'eel massive quasiparticles -- is orders of magnitude in nm \cite{pair-coherence-length}. 

To end this conclusions, it is in progress the proof, analogously to the relativistic massive case \cite{brown-ravenhall-operator}, if whether the attractive $s$-wave scattering potential can lead to bound states, that is, if the potential (\ref{Vs}), provided $g^2>e^2$, could favour $s$-wave massless bipolarons. The possible emergence of such a kind of Cooper-type electron-polaron--electron-polaron (hole-polaron--hole-polaron) condensate draws attention to superconductivity in graphene \cite{graphene-superconductor}.  
   
\subsection*{Acknowledgments} 
The authors thank J.A. Helay\"el-Neto and D.H.T. Franco, as well as to the anonymous referee for helpful comments and suggestions. O.M.D.C. dedicates this work to his father (Oswaldo Del Cima, {\it in memoriam}), mother (Victoria M. Del Cima, {\it in memoriam}), daughter (Vittoria) and son (Enzo). In spite of the heavy budget cuts on science, scientific research and education promoted this year by the Brazilian Ministry of Education (MEC), CAPES-Brazil is acknowledged for invaluable financial help. FAPEMIG is unworthy owing to the extinction of the scientific initiation fellowship program for undergraduate students executed by the Minas Gerais State government.

\vspace{0.5cm}
\subsection*{Author contribution statement} 
All authors have been contributed equally.

\end{document}